\documentclass{svjour3}                     \smartqed  
\usepackage[pdftex]{graphicx}

\usepackage{threeparttable}
\usepackage{footnote}
\usepackage{setspace}

\newcounter{num}

\renewcommand{\thetable}{\Roman{table}}
\renewcommand{\figurename}{Fig.}
\renewcommand{\thefigure}{\arabic{figure}}

\newcommand{\bhline}[1]{\noalign{\hrule height #1}}  

\usepackage{newtxtext,newtxmath}
\fontsize{10pt}{18pt}\selectfont

\begin{document}

\title{Infrared Absorption and its Sources of CdZnTe at Cryogenic Temperature
}

\titlerunning{Infrared Absorption and its Sources of CdZnTe at Cryogenic Temperature}        

\author{Hiroshi Maeshima$^{1,2,*}$, Kosei Matsumoto$^{1,2}$, Yasuhiro Hirahara$^{3}$, Takao Nakagawa$^{2}$, Ryoichi Koga$^{3}$, Yusuke Hanamura$^{3}$, Takehiko Wada$^{2}$ , Koichi Nagase$^{2}$, Shinki Oyabu$^{4}$, Toyoaki Suzuki$^{2}$, Takuma Kokusho$^{5}$, Hidehiro Kaneda$^{5}$, and Daichi Ishikawa$^{5}$}

\authorrunning{H. Maeshima et al.} 

\institute{
1--Graduate School of Science, The University of Tokyo, 7-3-1 Hongo, Bunkyo-ku, Tokyo 113-0033, Japan.\\ 
2--Institute of Space and Astronautical Science, Japan Aerospace Exploration Agency, 3-1-1 Yoshinodai, Chuo-ku, Sagamihara, Kanagawa 252-5210, Japan.   \\  
3--Graduate School of Environmental Studies, Nagoya University,  Furo-cho, Chikusa-ku, Nagoya, Aichi 464-8601, Japan. \\ 
4--Institute of Liberal Arts and Sciences, Tokushima University, 1-1 Minami-Jyosanjima, Tokushima-shi, Tokuhsima 770-8502, Japan.\\
5--Graduate School of Science, Nagoya University, Furo-cho, Chikusa-ku, Nagoya, Aichi 464-8602, Japan. \\
*--\email{maeshima@ir.isas.jaxa.jp} \\
\quad Telephone: +81-50-336-23909 \\
\quad FAX: +81-42-786-7202 
}


\maketitle

\newpage

\begin{abstract} 
To reveal the infrared absorption causes in the wavelength region between electronic and lattice absorptions, we measured the temperature dependence of the absorption coefficient of $p$-type low-resistivity ($\sim 10^2~{\rm \Omega cm}$) CdZnTe crystals. 
We measured the absorption coefficients of CdZnTe crystals in four-wavelength bands ($\lambda=6.45$, 10.6, 11.6, 15.1$~\mu$m) over the temperature range of $T=8.6$--300 K with an originally developed system. 
The CdZnTe absorption coefficient was measured to be  $\alpha=0.3$--0.5 ${\rm cm^{-1}}$ at $T=300$ K and $\alpha=0.4$--0.9 ${\rm cm^{-1}}$ at $T=8.6$ K in the investigated wavelength range. 
With an absorption model based on transitions of free holes and holes trapped at an acceptor level, we conclude that the  absorption due to free holes at $T=150$--300 K and that due to trapped-holes at $T<50$ K are dominant absorption causes in CdZnTe. 
We also discuss a method to predict the CdZnTe absorption coefficient at cryogenic temperature based on the room-temperature resistivity. 
\keywords{CdZnTe \and infrared transmittance \and cryogenic material \and absorption coefficient}
\end{abstract} 
\section*{INTRODUCTION}\label{sec:intro}

Cadmium zinc telluride (CdZnTe) is a compound semiconductor of the II-VI type, commonly used as X-ray detectors \cite{Eisen1998} or substrates for the growth of epitaxial layers of mercury cadmium telluride (MCT) for infrared detector arrays \cite{Norton2002}. 
Large-size single-crystal CdZnTe growth techniques have been developed for such applications. 
Single-crystal ingots of CdZnTe with 5-inch diameter are now commercially available, with 6-inch crystals in the experimental stage \cite{Noda2011}.  

CdZnTe is also promising as an infrared optical material with wavelengths between 5--20 $\mu$m. 
In general, the electronic and lattice absorptions characterize intrinsic absorption, which is the immutable absorption even for pure semiconductors \cite{Deutsch1975}.  
CdZnTe has little intrinsic absorption in the infrared wavelength range.
Also, CdZnTe has extrinsic absorption, such as free-carrier absorption due to impurities and attenuation due to Te precipitates \cite{Noda2011,Sarugaku2017}. 
As a result, we anticipate that CdZnTe can be used as an infrared optical material by  controlling its extrinsic properties (e.g., impurity or Te precipitates). 

In addition, these infrared materials are frequently used in low-temperature environments to reduce thermal background radiation, particularly in astronomical observation optics (e.g., SPICA Mid-Infrared Instrument \cite{Wada2020} on board the SPICA mission\footnote{SPICA had been one of the candidates for the fifth M-class mission in the ESA Cosmic Vision but was canceled on financial grounds in October 2020. } \cite{Nakagawa2020,Roelfsema2018}). 
As a result, the ability of infrared materials to transmit light at low temperatures is essential for astronomical applications.

An immersion grating is an optical material application of CdZnTe \cite{ikeda2015a}. 
Immersion gratings are diffraction gratings with a  grooved surface immersed in a high-refractive-index material \cite{Leitner1975,Marsh2007}. 
CdZnTe is an attractive material for  immersion grating at wavelengths of $\lambda=$10--20 $\mu$m due to its high-refractive index  ($n\sim 2.7$) and transparency \cite{Sukegawa2012,Kaji2014}. 
Owing to its high-refractive index $n$, an immersion grating can be downsized $1/n$-sized in length compared to conventional gratings. 
In size-constrained situations, the $1/n$-downsizing of a grating has a significant impact on the design of a compact optical system (e.g., astronomical satellites, systems in cryostat dewars). 
CdZnTe-immersion grating is expected to be applied in space telescopes \cite{Sarugaku2012}.

The room-temperature absorption coefficient in infrared wavelengths is known to be  resistivity dependent. 
Kaji et al. \cite{Kaji2014} investigated the absorption coefficient of CdZnTe crystals with  resistivities of $\rho=3.5\times 10^{10}~{\rm \Omega cm}$ and $\rho\sim 1\times 10^2~{\rm \Omega cm}$ at room temperature. 
They discovered that the absorption coefficient of low-resistivity CdZnTe was higher than that of high-resistivity CdZnTe and increased with wavelength at $\lambda=$ 5--20 $\mu$m.  
This high absorption is considered to be caused by free-carrier absorption \cite{Deutsch1975,Kaji2014}.  
When low-resistivity CdZnTe is cooled down to cryogenic temperature, the free-carrier absorption coefficient becomes smaller than the coefficient at room temperature due to free-carriers freeze-out.  
Therefore, low-resistivity CdZnTe, which is more readily available than high-resistivity CdZnTe, is expected to be suitable for cryogenic infrared materials. 
However, because few previous studies on the optical performance of CdZnTe at low temperatures, absorption causes in CdZnTe at cryogenic temperature are unknown. 
The optical performance of low-resistivity CdZnTe must be studied in order to develop cryogenic infrared materials (e.g., immersion gratings \cite{ikeda2015a}). 

In this paper, we investigate the process for mid-infrared photon absorption by measuring the temperature dependence of the absorption coefficient of CdZnTe in the temperature range of $T=8.6$--300 K. 
We expect cryogenic absorption coefficients of CdZnTe crystals to be predictable from room-temperature physical properties after revealing the possible processes for the mid-infrared photon absorption. 
 
\section*{METHOD}

This section describes the method used to measure the absorption coefficient.  
First, we describe the derivation method of the absorption coefficient from transmittance measurements. 
Taking into account the multiple reflections, we express the total transmittance $\tau$ as follows:
\begin{eqnarray}
\tau=\frac{(1-R)^2\exp(-\alpha t)}{1-R^2\exp(-2\alpha t)},
\end{eqnarray}
where $\alpha$ is the absorption coefficient, $R=(n-1)^2/(n+1)^2$ is the Fresnel surface reflectivity, and $t$ is the sample thickness \cite{Sarugaku2017}. 
We can estimate the absorption coefficient by measuring the transmittances ($\tau_1,~\tau_2$) of samples made of the same material but with different thicknesses ($t_1,~t_2$) as follows:
\begin{eqnarray}
\alpha=-\frac{\ln (\tau_1/\tau_2)}{t_1-t_2}+X,\\
X=\frac{1}{t_1-t_2}\ln\left[\frac{1-R^2\exp(-2\alpha t_2)}{1-R^2\exp(-2\alpha t_1)}\right].
\end{eqnarray}
In this case, we assume the common surface reflectivity $R$. 
The term $X$ arises from the effect of the multiple reflections. 
In the case of the refractive index $n=2.7$, the $X/\alpha$ is less than 0.1 at any $\alpha$ value. 
If $X$ is included in the systematic error, $\alpha$ can be approximated as follows:
\begin{eqnarray}
\alpha=-\frac{\ln (\tau_1/\tau_2)}{t_1-t_2}. 
\end{eqnarray}
Therefore, $\alpha$ can be estimated without assuming the $R$ value. 
In this study, we use Eq. 4 to calculate $\alpha$ from the measured transmittance $\tau$. 

Table \ref{tab:1} shows the specifications of the CdZnTe samples used in this study. 
JX Nippon Mining \& Metals Corporation produced the ${\rm Cd_{0.96}Zn_{0.04}Te}$ single crystals used in the measurements.  
The vertical gradient freezing (VGF) method is used to make the CdZnTe ingot. 
In the main text, the conductivity is controlled to $p$-type for the ingot and $n$-type for Appendix A. 
The substrates were then cut from the ingot.  
To confirm whether the predicted change in absorption coefficient with temperature exists, we controlled the resistivity to the same order ($\rho\sim 10^{2}~{\rm \Omega cm}$) as the low-resistivity CdZnTe measured by Sarugaku et al. \cite{Sarugaku2017}. 
The ingot was then annealed to reduce scattering caused by Te precipitates by reducing precipitate size \cite{Noda2011}. 
The incident and exit surfaces are polished with a surface roughness of $Ra<$1 nm to reduce scattering loss on the surface, where $Ra$ is the calculated average roughness. 
To derive the absorption coefficient from the thickness dependence of the transmittance, samples with two types of thicknesses, lowR-t1 and lowR-t10, are taken from the same ingot. 

 \begin{table*}
\caption{Properties of the CdZnTe single crystals. The specifications are measured at room temperature. ($Ra$: arithmetic average roughness of the surface) }
\label{tab:1}       \begin{tabular}{ccccccc}
\hline\noalign{\smallskip}
sample 	& incident-surface size 		& thickness  	& resistivity 				& conductivity 	& Te particle size	&	$Ra$ \\
		&[mm$^2$]			&  [mm] 		& [${\rm \Omega}$cm] 	&  type			&  [$\mu$m] 		&	[nm] \\
\noalign{\smallskip}\bhline{1.5pt} \noalign{\smallskip}
lowR-t1 	&7$\times$7 		& 1.01 		& (0.5--1.3)$\times 10^2$	& $p$ 			& $<2$			& $<1$ \\
lowR-t10	&7$\times$7 		& 9.91		& (0.5--1.3)$\times 10^2$	& $p$			& $<2$			& $<1$\\
\noalign{\smallskip}\hline
\end{tabular}
\end{table*}

Generally, infrared transmittance is measured using a Fourier transform spectrometer (FTS).  
However, commercial FTS products have numerous constraints, making it practically difficult to adjust the position of the sample cooling stage.
In addition, for thick and high-refractive-index ($n$) substrates such as CdZnTe, transmittance measurement with a typical uncollimated FTS may result in significant systematic uncertainty due to the defocus effect \cite{Kaji2014}. 
Installing a thick and high-$n$ substrate in the sample compartment would lengthen the optical path and shift the detector's focus position backward. 
Furthermore, the focal position is wavelength dependent due to the wavelength dependence of the refractive index $n$. 
Because of this, defocus large systematic errors in transmittance measurement would occur.

We built an original measurement system on an optical bench using a collimated beam and a cryostat to solve the adjustment difficulty and the defocus problem. 
We anticipate that the defocus effect will not be an issue by irradiating the collimated beam with the sample \cite{Kaji2014}. 
Figure 1 shows the originally developed measurement system and Table \ref{tab:2} summarizes the system's specifications.   
To irradiate samples, we use a globar lamp (1 in Fig. 1; SLS303, Throlabs Japan Inc.) as an infrared light source.
The lamp's irradiated beam is guided by a polycrystalline infrared fiber (3) and is recollimated with a lens (4). 
We can change the wavelength of the collimated beam by mounting four bandpass filters ($\lambda=6.45,~ 10.6,~ 11.6,~ 15.1~\mu$m; listed in Table \ref{tab:2}) in the filter wheel. 
To pass through the sample-holder aperture ($\phi 5$ mm), the infrared light beam is adjusted to $\phi 2$ mm with an aperture (7). 
To monitor the time variation, we split the incident beam by the beam splitter (8), and the reflected beam B is focused on a Peltier-cooled MCT detector (13; PVMI-4TE-10.6-1x1-TO8-wZnSeAR-35; VIGO system S.A.). 
Beam A is transmitted to the sample (11) with the incident angle of $\sim 0^\circ$ and focused on another liquid-nitrogen (${\rm LN_2}$) cooled MCT detector (17: IOH-1064, Bomem inc.) to measure the transmitted-light power. 
We use an optical chopper (5; model 300 CD, Scitec instruments) with a frequency of $\sim 1$ kHz  and lock-in amplifiers (SR830, Standard Research Systems) connected to the detectors because the incident beam to the sample is  relatively faint ($\sim 10^{-5}~{\rm W~cm^{-2}}$ in the 10.6 $\mu$m band right after the aperture 7).

\begin{figure*}
	\includegraphics[width=129mm]{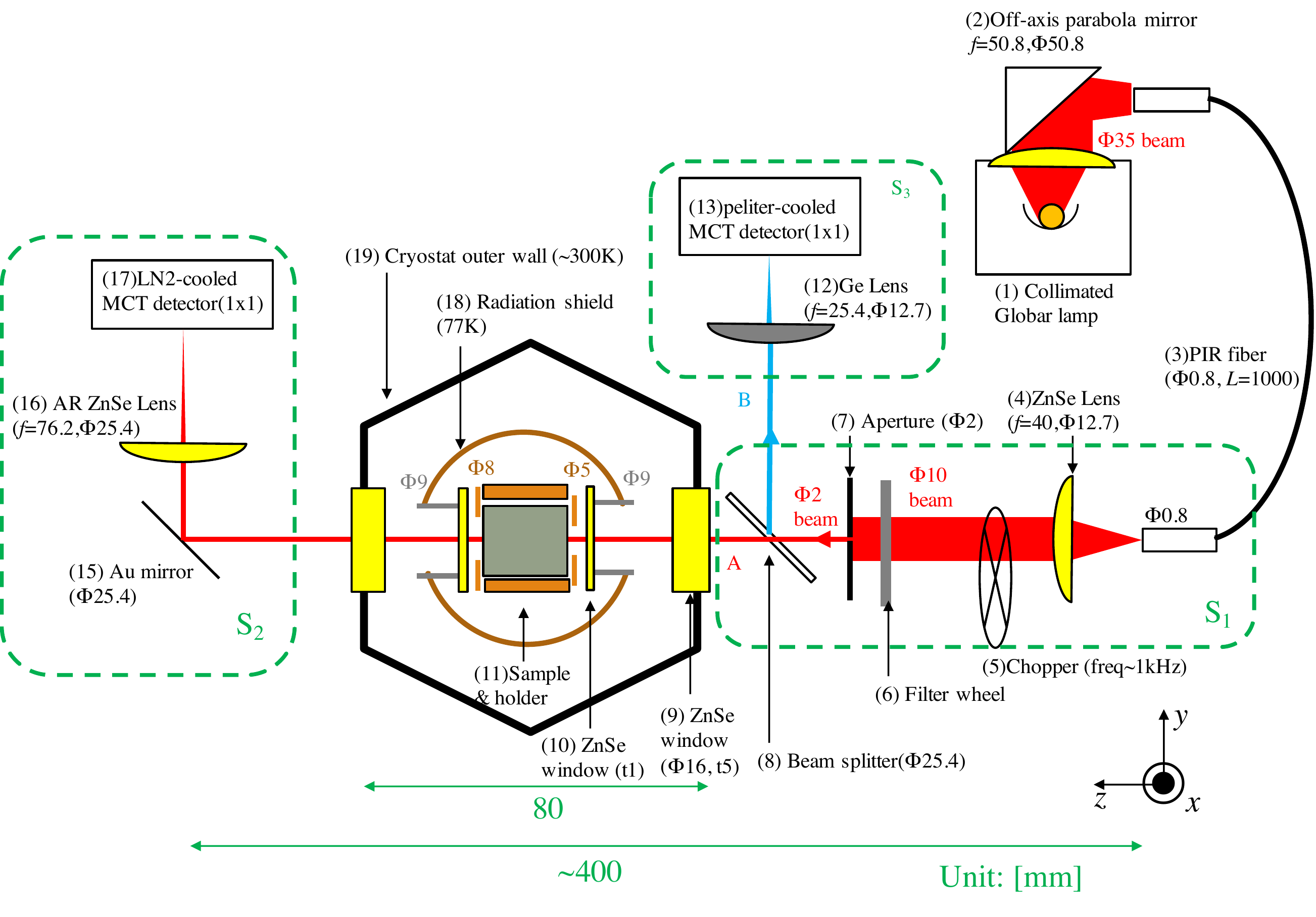}
	\caption{The measurement system. 
Each component is described in the text. 
Red and blue lines show the beam path. 
Dimensions are in millimeters. 
Focal length $f$ is indicated for each lens. 
Thickness, diameter, length of each component are denoted by $t$, $\Phi$, and $L$, respectively. 
Components in green dashed areas are mounted on each stage plate. 
The stages $S_1$, $S_2$ $S_3$  can move in two directions vertical to the beam axis. 
The temperatures of the cryostat outer wall (19) and the radiation shield (18) are $\sim 300$ K and $\sim 77$ K, respectively, when a sample is cooled to the lowest temperature (8.6 K). 
Details of sample holder structure are described in Appendix B. 
}
	\label{fig:1}
\end{figure*}

\begin{table*}[h]
\caption{Specifications of the measurement system. \label{tab:2}}
\begin{tabular}{lll} \hline 
Parameters				& Specifications					& Notes	\\ \bhline{1.5pt}
Light source				& Silicon Nitride Globar  & \\
Detectors				& MCT detectors	&  LN$_2$-cooled/Peltier-cooled \\
Beam size				& $\phi$2 mm			& \\
Temperature 				& $T=$8.6--300 K		& see results section \\
incident angle 			& $\sim 0^\circ$	\\
Bandcenter $\pm$ HWHM  	& $\lambda=6.45\pm0.05,~10.6\pm0.8, $ 	& \\
						& $11.6\pm 0.4,~15.1\pm0.6$ ${\rm \mu m}$ 		& \\ \hline
\end{tabular}
\end{table*}

We use a cryostat with a two-stage Gifford-Mcmahon cryocooler (PS24SS; Nagase \& Co., Ltd) to cool a sample to cryogenic temperature.
Furthermore, we attach anti-reflective-coated (AR-coated) ZnSe windows with thicknesses of 5 mm and 1 mm to the vacuum chamber and the radiation shield in the chamber, respectively, to transmit the mid-infrared beam. 
To avoid thermal conduction, we reduce the gas pressure inside the chamber  to $<10^{-2}$ Pa. 
A built-in calibrated Si-diode sensor measures the temperature of the cold head.
To stabilize the temperature during measurement, we install a heater near the cold head. 
We prepare two types of holders to cool samples: one for the thick sample and the other for the thin sample.
The details of the holders are described in Appendix B. 

Next, we describe the measurement procedure. 
A sample's transmittance $\tau$ is calculated as
\begin{eqnarray}
\tau(\lambda,T)=\frac{V_{\rm A,smp}(\lambda,T)/V_{\rm B,smp}(\lambda,T)}{V_{\rm A,ref}(\lambda)/V_{\rm B,ref}(\lambda)} \label{eq:3},
\end{eqnarray}
where $V_{x,{\rm smp}}~(x=A $ or $B)$ is the power detected on the detector of the beam $x$ side with the sample and $V_{x,{\rm  ref}}$ is the power detected without the sample.  
To reduce the effect of time variation in incident light powers, we measure powers $V_{\rm A, smp}$ and $V_{\rm B, smp}$ simultaneously, as well as $V_{\rm A, ref}$ and $V_{\rm B,ref}$.  
We measure the $V_{\rm A,ref}(\lambda)/V_{\rm B,ref}(\lambda)$ at 300 K because the denominator $V_{\rm A,ref}/V_{\rm B,ref}$ is a constant value against temperature change.  
We measure the transmittance of two samples at the four bands described in Table \ref{tab:2}. 

We cool the samples at a rate of $<6$ K min$^{-1}$ to reduce thermal stress on the sample. 
During the transmittance measurement, the temperature of the cold head is controlled within $\pm 0.1$ K by using a heater controller.
Because thermal shrink affects the beam vignetting, it is necessary to optimize the beam alignment  during the measurement. 
Also, because sample's position shifts by $\sim 1$ mm due to thermal shrinkage of the cold-head shaft, the beam power measured at the aperture may be vignetted at $\sim 3\%$. 
Thus, at each measurement temperature, we adjust the stages $S_1$, $S_2$, and $S_3$ (see Fig. 1) in two vertical directions to the optical axis to allow the beam to pass through the center of the holder aperture.  
Subsequently, because the beam is collimated, the adjustment does not affect the focal length. 
As a result, we can adjust the beam position by adjusting the three stages.
We measure the transmittance of thin and thick samples at various temperatures ranging from about $T=300$ K to the lowest temperature ($8.6$ K; see the following section).

\section*{RESULTS}
Figure 2 shows the absorption coefficients $\alpha$ of the CdZnTe at $\lambda=6.45$, 10.6, 11.6, and 15.1 $\mu$m from room temperature to a cryogenic temperature at $T=8.6$ K. 

We calibrate the systematic temperature error caused by the Seebeck effect by averaging the measured temperatures with the voltage polarity swapped.
Based on the measurement system's specification accuracy (model 218, Lakeshore inc.), we estimate the temperature uncertainty as $\Delta T=10,~5,~2,~1,~0.3,~ 0.05~$K at $T\sim 300,~200,~100,~50,~8.6$ K, respectively.  

The uncertainty of the measured absorption coefficient is mainly caused by the reproducibility of optimizing the beam alignment described in the previous section. 
We hence assume that the reproducibility is the same for all cases of temperatures, wavelengths bands, and samples because we perform the stage adjustments following a similar approach.  
We measure the 1$\sigma$ reproducibility of the transmittance $\Delta\tau/\tau=0.011$ at 300 K and apply it to all the measured $\tau$. 
From the error propagation law using Eq. 4, the 1$\sigma$ uncertainty of $\alpha$ is estimated as $\Delta\alpha=0.016 ~{\rm cm^{-1}}$.

As shown in Fig. 2, the measured $\alpha$ ranges from 0.3--0.5 ${\rm cm^{-1}}$ at $T=300$ K and from 0.4--0.9 ${\rm cm^{-1}}$ at $T=8.6$ K. 
At each temperature, the absorption coefficient $\alpha$ at the longer wavelength is larger than the value at the shorter wavelength.
At $T=100$--300 K, the $\alpha (\lambda=6.45~{\rm \mu m})$ decreases with cooling, whereas the $\alpha(\lambda=15.1~{\rm \mu m})$ increases. 
In the temperature range of $T=100$--300 K, the temperature dependences of $\alpha(\lambda=10.6~{\rm \mu m})$ and $\alpha(\lambda=11.6~{\rm \mu m})$ are small compared to the uncertainty. 
At all measured wavelengths, the absorption coefficient $\alpha$ increases by 16--40\% from $T=100$ K to $T=50$ K. 
At $T<50$ K, the temperature dependence of the  $\alpha$ becomes small.  
The dependence of the absorption coefficients on temperature is discussed in the next section.

The result is then compared to the free-carrier expectation \cite{Sarugaku2017}.
The absorption rate caused by free carriers is proportional to the free carrier's density. 
At low temperatures, the carrier density is proportional to $f(T)=\exp(-E_a/2k_BT)$, where $E_a$ is the energy separation of the donor or acceptor level  from the conduction or valence band \cite{kittel1976} and $k_B$ is the Boltzmann constant. 
If we assume that $E_a$ of CdZnTe is consistent with that of CdTe  ($E_a=60$ meV; \cite{Ahmad2015}) for simplicity, cooling from $T=300$ K to $T<50$ K reduces the free-carrier density by less than $f(T=50~{\rm K})/f(T=300~{\rm K}) \sim 0.003$.   
The proportional relationship would also reduce the free-carrier absorption coefficient by less than $\sim 0.003$ times. 
Even for the case of multi-impurity levels, free-carriers are estimated to be frozen out because the reported acceptor levels of CdTe are $\sim 60$ meV or $\sim 150$ meV \cite{Ahmad2015,molva1984}.  
In contrast to the free-carrier expectation, the measured $\alpha$ increases by about 1.3--1.8 times from $T=300$ K to $T=8.6$ K \cite{Sarugaku2017}. 
In the following section, we discuss this apparent inconsistency.

\begin{figure*}
	\includegraphics[width=84mm]{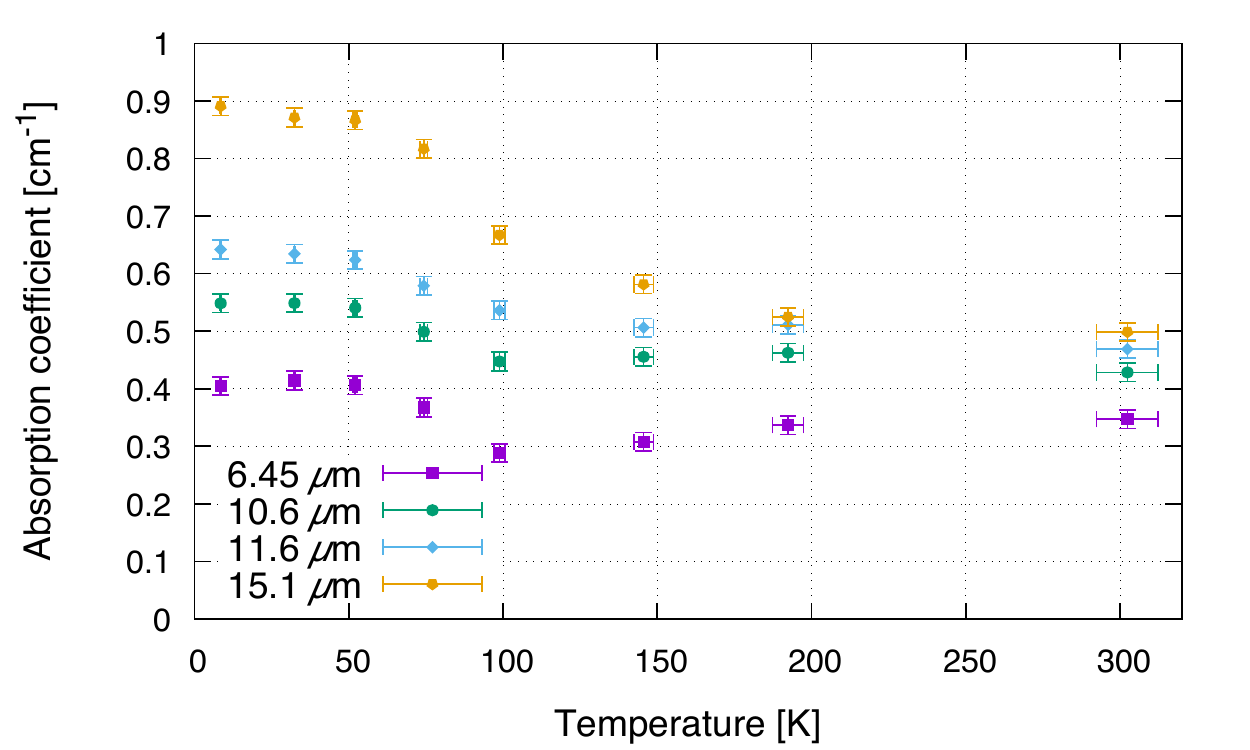}
	\caption{The absorption coefficient of the CdZnTe at the wavelength of $\lambda=6.45$ (purple), 10.6 (green), 11.6 (sky blue),  and 15.1 (orange) $\mu$m. 
Vertical error bars show the 1$\sigma$ uncertainty range of the absorption coefficient. 
Horizontal error bars show the accuracy of the temperature. 
}
	\label{fig:alpha}
\end{figure*} 
\section*{DISCUSSION}
In this section, we discuss absorption causes in CdZnTe. 
As shown in the previous section, our result reveals that the $\alpha$ increases at the cryogenic temperature, contrary to the free-carrier expectation. 
This disparity suggests that, in addition to free-carrier absorption, other causes influence the absorption coefficient. 

\begin{figure*}
	\includegraphics[width=84mm]{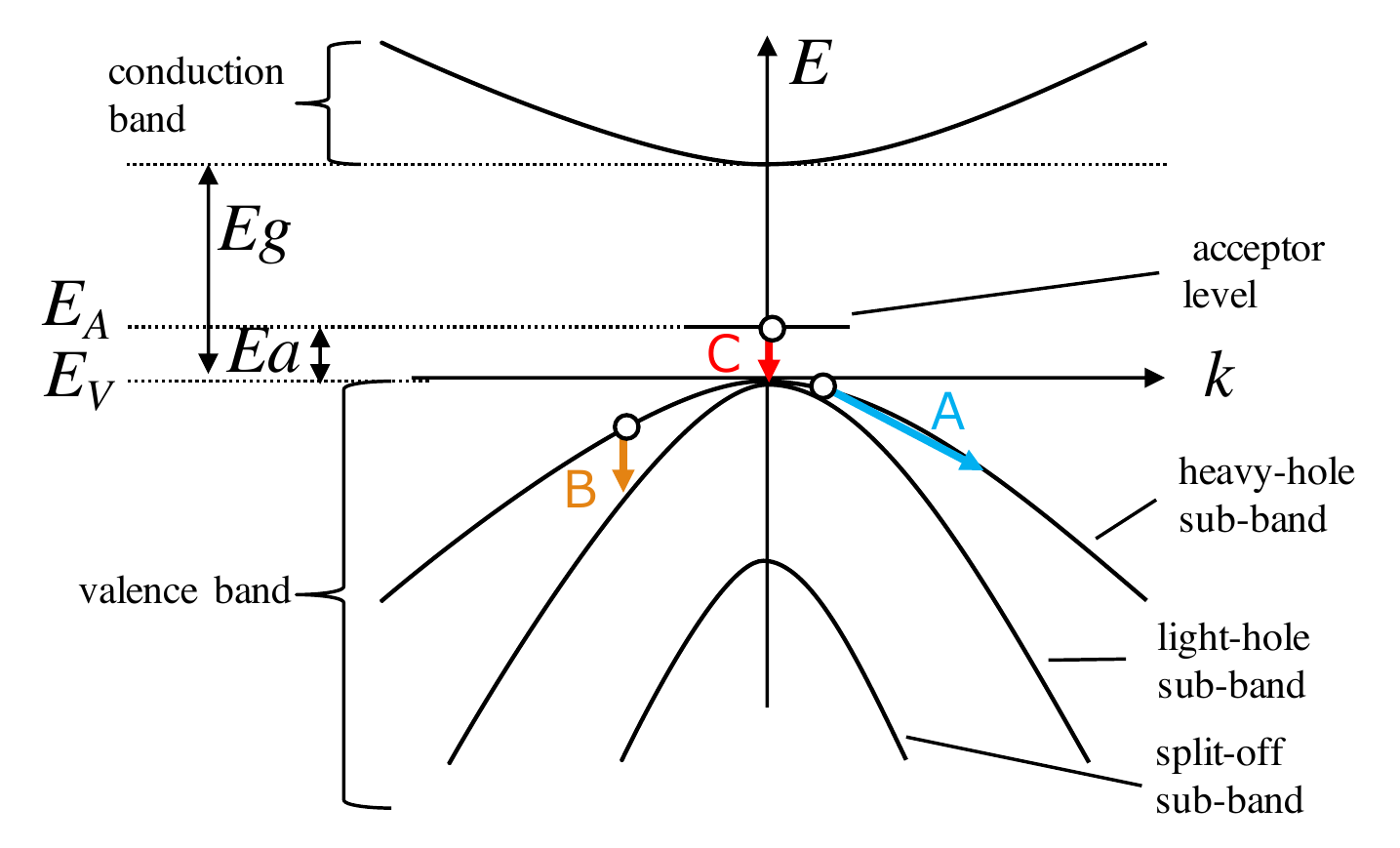}
	\caption{Schematic view of the band structure of CdZnTe. 
	The horizontal axis shows wave vector $k$ and the vertical one shows energy $E$. 
	The valence band is separated into three sub-bands: the heavy-hole, light-hole, and split-off sub-bands.
	The energy levels of the top valence band and the acceptor are denoted by $E_V$ and $E_A$, respectively. 
	The band-gap energy is denoted by $E_g$ and the gap energy between acceptor level and the valence band is $E_a$. 
	The transitions A, B, and C are the absorption transitions of the model described in the text.}
	\label{fig:4}
\end{figure*}

\subsection*{\bf Absorption model}
To discuss the dominant absorption cause, we build a model of the $\alpha$ of the CdZnTe.  
To discuss the causes at room temperature, we include the absorption causes considered in the CdZnTe study at room temperature \cite{Sarugaku2017} in our model.
Figure 3 shows a schematic view of the band structure of $p$-type CdZnTe, and hole transitions related to absorption in the model. 
At room temperature, $p$-type CdZnTe  absorbs due to two types of free-carrier transitions: (A)  transitions within the same valence sub-band, (B)  transitions from the heavy-hole sub-band to the light-hole sub-band (e.g., \cite{Basu2003}). 
According to the room-temperature study, CdZnTe has attenuation due to Te precipitates and the lattice absorption \cite{Sarugaku2017}.

The absorption causes in CdZnTe are well discussed at room temperature \cite{Sarugaku2017}, but they are not studied at low temperatures. 
As a result, we include the absorption causes considered in a previous cryogenic CdTe study \cite{Capek1973} to discuss the temperature dependence of the absorption coefficient. 
In this case, we assume that the physical properties of CdZnTe are similar to those of CdTe due to the low Zn abundance ratio ($\sim 4\%$). 
\v{C}\'{a}pek et al. \cite{Capek1973} showed the existence of absorption of  the hole transition from the acceptor level to the valence band (C in Fig. 3). 

The total absorption coefficient $\alpha_{\rm model}$ is expressed as 
\begin{eqnarray}
\alpha_{\rm model} &=&  N_{\rm free}(\sigma_{\rm intra}  +\sigma_{\rm inter} ) \nonumber\\
&&+N_{\rm trapped} \sigma_{\rm trapped} +\alpha_{\rm Te} +\alpha_{\rm lattice},
\end{eqnarray}
where $\sigma_{\rm intra}$, $\sigma_{\rm inter}$, and $\sigma_{\rm trapped}$ are the absorption cross-sections of the (A) intraband, (B) interband, and (C) acceptor-valence band transition, respectively. 
The density of free holes is denoted by $N_{\rm free}$, while the density of trapped holes is denoted by $N_{\rm trapped}$. 
Attenuation due to Te precipitates is denoted by $\alpha_{\rm Te}$, and the lattice absorption is denoted by $\alpha_{\rm lattice}$.

The Drude model (e.g., \cite{Kudo1996}) is used to calculate the cross-section of (A) intraband absorption $\sigma_{\rm intra}$. 
The absorption coefficient in the Drude model can be calculated using the equation of motion of a free carrier with an attenuation term. 
We derive the $\sigma_{\rm intra}$ as
\begin{eqnarray}
\sigma_{\rm intra}	&=&	\frac{q^2}{\pi n c^5 m^* \tau_{R}(T)\tilde{\nu}^2} \nonumber \\
				&=&	b_0\left(\frac{\tilde{\nu}}{943~{\rm cm^{-1}}}\right)^{-2}\left(\frac{T}{300 ~{\rm K} }\right)^{-1.5},
\end{eqnarray}
where $b_0$ is a proportional factor, $\tilde{\nu}$ is the wavenumber ($\tilde{\nu}=1/\lambda$), $q$ is the effective charge, $n$ is the refractive index, $c$ is the light speed, $m^*$ is the effective mass, and $\tau_R$ is the relaxation time. 
The relaxation time $\tau_R$ is the mean time interval of free-hole's collisions with main scatterers. 
We assume that $\tau_R$ is proportional to $T^{-1.5}$ because lattice vibrations at  $T=100$--300 K primarily caused the carrier scattering for CdTe case \cite{Ahmad2015}. 
Although the temperature dependence of hole mobility is not proportional to $T^{-1.5}$ at $T<100$ K, we assume that, at such low temperatures, intraband absorption becomes negligible because of free-hole freeze-out.
Therefore, we apply the single power-law function as the $\tau_R(T)$ in the model. 

Next, we describe the cross-section of (B) interband absorption $\sigma_{\rm inter}$. 
We consider the absorption cross-section of the direct transition of holes from the heavy-hole sub-band to the light-hole sub-band.
In this case, we assume that the bands have spherical energy surfaces with a parabolic dispersion relation and that the occupational probability at each band has a Maxwellian energy distribution.
The cross-section $\sigma_{\rm inter}$ is estimated in a similar approach as general $p$-type semiconductors described in Basu \cite{Basu2003} as follows:
\begin{eqnarray}
\sigma_{\rm inter}&=&a_0\left(\frac{\tilde{\nu}}{943\;{\rm cm^{-1}}}\right)^{0.5}\left(\frac{T}{300{\rm\; K}}\right)^{-1.5} \nonumber\\
&&\times \left[1-\exp\left(-\frac{hc\tilde{\nu}}{k_BT}\right)\right]\exp\left(-a_1 \frac{hc\tilde{\nu}}{k_BT}\right), \\
a_1&=&\left(\frac{m_h}{m_l}-1\right)^{-1},
\end{eqnarray}
where $a_0$ is a proportional factor, $m_h/m_l$ is the effective mass ratio of heavy and light holes, and $h$ is the Planck constant.

We discuss the cross-section of (C) trapped-hole absorption $\sigma_{\rm trapped}$. 
\v{C}\'{a}pek et al. \cite{Capek1973} measured the $\alpha$ of CdTe at cryogenic temperature and concluded that the trapped-hole absorption exists.
We assume that the cross-section $\sigma_{\rm trapped}$ is a function of wavelength but not of temperature. 
This assumption is made because the temperature dependence of the absorption cross-section is primarily due to the temperature dependence of the occupational probability difference between initial acceptor state-$i$  and final state-$j$ in the valence band ($f_i-f_j$) \cite{Basu2003}. 
At low temperatures in the acceptor state, the majority of the holes will be trapped, and free holes will be exhausted. 
Because trapped-hole absorption will be dominant at cryogenic temperatures, we assume that  $f_i\sim 1$, $f_j\sim 0$ and trapped-hole cross-section is temperature independent.
 
We then describe the densities of free holes $N_{\rm free}$ and trapped holes $N_{\rm trapped}$. 
There are two possible types of free holes. 
One is "intrinsic" free holes, which are released by the transition from the conduction band to the valence band,  and the other is "extrinsic" free holes, which are released by the transition from an acceptor band to the valence band.
The total free-hole density at the room temperature can be estimated with the resistivity $\rho=50$--$130~{\rm \Omega cm}$ as
\begin{eqnarray}
N_{\rm free}(300~{\rm K})=\frac{1}{q\rho\mu}=(9\pm 5)\times 10^{14}~{\rm cm^{-3}},
\end{eqnarray}
where $q$ is the charge of a single hole. 
We assume that the hole mobility $\mu$ is the same as that of CdTe ($\mu=80~{\rm cm^{2}V^{-1}sec^{-1}}$, \cite{Ahmad2015}) because of the low percentage of Zn ($\sim 4\%$) in our CdZnTe samples. 
On the other hand, the intrinsic hole density at the room temperature is estimated as
\begin{eqnarray}
N_{\rm free,intrinsic}(300~{\rm  K})&=&2\left(\frac{k_BT}{2\pi \hbar^2}\right)^{3/2}(m_{e*}m_*)^{3/4}\exp\left(-\frac{E_g}{2k_BT}\right)\nonumber\\
				&\sim&  10^{6}~{\rm cm^{-3}} \label{eq:n_intrinsic},
\end{eqnarray}
where $m_*$ is the effective hole mass ($m_*/m_e=0.30$ for CdTe; \cite{wakaki2007}), $m_{e*}$ is the effective electron mass ($m_{e*}/m_e=0.11$ for CdTe; \cite{wakaki2007}), and $E_g$ is the bandgap energy between the valence band and the conduction band ($E_g\sim 1.5$ eV \cite{Noda2011}) \cite{kittel1976}. 
The free holes at the room temperature are dominated by the extrinsic free holes released from the acceptor level, and we ignore the intrinsic free holes. 
At lower temperatures, the ratio of the extrinsic-free-hole density to the intrinsic-free-hole density becomes larger since the intrinsic energy gap is larger than that between the acceptor band and the valence band. 
Therefore, we also ignore the intrinsic free holes at lower temperatures. 

We derive the temperature dependence of the $N_{\rm free}$ similarly to that of Ahmad \cite{Ahmad2015}. 
We obtain an analytical equation of the free-hole density by coupling the following three equations: the equation for the density of holes trapped at the acceptor level, the equation for free-hole density based on the effective density of states, and the conservation equation of the sum of the free and trapped-hole densities.  
We derive the $N_{\rm free}$ as
\begin{eqnarray}
N_{\rm free} &=& N_A\frac{2}{\sqrt{1+8(N_A/N_V)\exp(E_a/k_BT)}+1}, \label{eq:nfree}
\end{eqnarray} 
where $N_A$ is the acceptor density, $N_V$ is the effective density of states, $E_a$ is the energy gap between the acceptor level and the valence band's top. 
In this case, we assume that there is a single acceptor level since Ahmad \cite{Ahmad2015}  explained the temperature dependence of the CdTe free-hole density measured by them with a single-acceptor-level model.  
We also assume that the effect of the donor is negligible. In this equation, the $N_{\rm free}$ is roughly constant in a relatively high-temperature range (namely saturation regime) due to full ionization of acceptors, in a relatively low-temperature range (namely the ionization regime),  whereas the $N_{\rm free}$ is proportional to $\exp(-E_a/2k_BT)$ \cite{Ahmad2015}. 
According to Kittel \cite{kittel1976}, the effective density of states $N_V$ is given by  
\begin{eqnarray}
N_V	&=&	2(2\pi m_*k_BT/h^2)^{3/2} \nonumber \\
	&=&	4.12\times 10^{18}\left(\frac{m_*}{0.3m_e}\right)^{3/2} \left(\frac{T}{300{\rm \;K}}\right)^{3/2} {\rm \; cm^{-3}}, 
\end{eqnarray}
where $m_*$ represents the effective hole mass ($m_*/m_e=0.30$ for CdTe; \cite{wakaki2007}). 
The acceptor is assumed to be fully ionized at $T\sim 300$ K \cite{Ahmad2015} and the free-carrier density is assumed to be almost the same as the acceptor density. 
Thus, the acceptor density is estimated as 
\begin{eqnarray}
N_A\sim N_{\rm free}(300{\rm ~K})=(9\pm 5)\times 10^{14}~{\rm cm^{-3}}. 
\end{eqnarray}
In this model, we assume the $N_A$ value as $N_A=9\times 10^{14}~{\rm cm^{-3}}$.  
We estimate the uncertainty of $N_A$ as $\pm 5\times 10^{14}~{\rm cm^{-3}}$ from the uncertainty of the room-temperature resistivity. 
The effect of the uncertainty of $N_A$ is discussed in the next subsection. 
Because the sum of the free-hole density $N_{\rm free}$ and the trapped-hole density $N_{\rm trapped}$ equals  the acceptor density $N_A$, the trapped-hole density $N_{\rm trapped}$ is expressed as
\begin{eqnarray}
N_{\rm trapped} &=& N_A-N_{\rm free}. \label{eq:ntrap}
\end{eqnarray} 

We then discuss how the attenuation due to Te precipitates $\alpha_{\rm Te}$ affects our model. 
Sarugaku et al. \cite{Sarugaku2017} evaluated Mie scattering attenuation due to Te precipitates as $\alpha_{\rm Te}\leq  0.01~{\rm cm^{-1}}$ at $\lambda>5~\mu$m. 
Subsequently, because Te precipitates size in our samples is limited to $\leq 2~\mu$m in the same way in their samples, we assume that the $\alpha_{\rm Te}$ in our samples is as small as  $\leq 0.01~{\rm cm^{-1}}$.  
Also, the scattering attenuation is unaffected by temperature. 
As a result, we ignore the $\alpha_{\rm Te}$ assuming it makes only a minor small contribution to the $\alpha_{\rm model}$.

In addition, we discuss the lattice absorption $\alpha_{\rm lattice}$. 
Based on the empirical exponential relationship, the $\alpha_{\rm lattice}$ at room temperature is less than $2\times 10^{-3}~{\rm cm^{-1}}$ at $\lambda<15.1~\mu$m, according to Sarugaku et al. \cite{Sarugaku2017}. 
The lattice absorption strength is proportional to the difference in phonon creation and annihilation probabilities.  
The strength of a single-phonon process is unaffected by temperature, whereas the strength of a multi-phonon process decreases with temperature \cite{Kudo1996}. 
Therefore, the $\alpha_{\rm lattice}$ is less than $2\times 10^{-3}~ {\rm cm^{-1}}$ at all temperatures and wavelengths measured.
Because the $\alpha_{\rm lattice}$ is negligible compared to the total absorption coefficient, we ignore it in our model.  

To summarize the above discussion, we set our absorption model as follows:
\begin{eqnarray}
\alpha_{\rm model,\tilde{\nu}} &=&  N_{\rm free}(T)(\sigma_{\rm intra,\tilde{\nu}} (T) +\sigma_{\rm inter,\tilde{\nu}} (T)) \nonumber\\
&&+N_{\rm trapped} (T)\sigma_{\rm trapped}.  
\end{eqnarray}
The free parameters are the acceptor energy ($E_a$), the effective mass ratio ($m_h/m_l$) and three proportional factors ($b_0,~a_0,~\sigma_{\rm trapped}$).

\subsection*{\bf Model-fittings and discussions}
This section discusses what causes absorption in the CdZnTe by fitting our results with the absorption model described above. 
Because the wavelength dependence of $\sigma_{\rm trapped}$ is unknown, we perform fitting individually in each wavelength band. 
Figure 4 shows the model fitting result in each band ($\lambda=6.45,~10.6,~11.6,~15.1~\mu$m). 
Each fitted parameter is listed in Table \ref{tab:fitT}. 
The $N_A$ value is fixed to $N_A=9\times 10^{14}~{\rm cm^{-3}}$ as discussed in the previous subsection. 
It is worth noting that, in the fitting in the 6.45 $\mu$m band, the $b_0$ value is fixed to $b_0=1\times 10^{-16}~{\rm cm^{2}}$ based on the best-fit value in the other wavelength bands to secure convergence of the fitting. 
This is because that intraband absorption becomes smaller in the shorter wavelength band due to its wavelength dependence and it is difficult to fit the parameter $b_0$ at the shortest wavelength 6.45 $\mu$m. 

Like the measured results, the model shows an increase in the absorption coefficient at low temperatures and reproduces the measured temperature dependence across the entire temperature range. 
Two physical parameters ($E_a,~m_h/m_l$) and two proportional factors ( $a_0,~b_0$) are consistent within 1$\sigma$ uncertainty range among each wavelength band case. 
This fitting result shows that, in the temperature range of $T=50$--150 K, the dominant holes change from the free holes to the trapped holes, and the dominant absorption causes changes from free holes (A, B) to trapped holes (C) with cooling.

 \begin{figure*}
	\includegraphics[width=174mm]{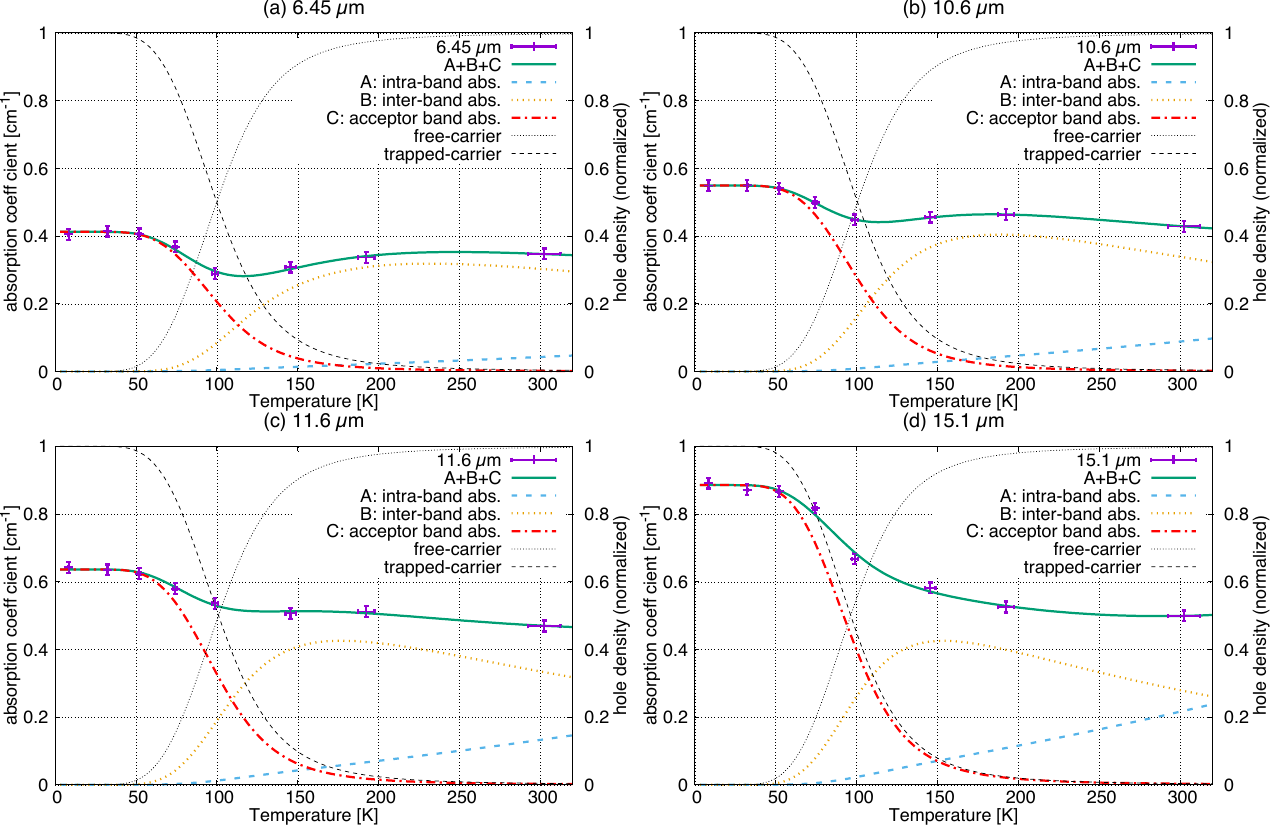}
	\caption{The temperature dependence of the absorption coefficient of the CdZnTe in the $\lambda=$ (a) 6.45, (b) 10.6, (c) 11.6, and (d) 15.1 $\mu$m bands and the fitted model. 
	 The purple points show the measured $\alpha$.
	 The error bars show 1$\sigma$ uncertainty ranges. 
	  Green solid lines show the total fitted absorption model. 
	  Sky-blue dashed, orange dotted, and red dashed-dotted lines show the fitted models of intraband, interband, and acceptor band absorption, respectively. 
	  The black dotted and dashed lines show the normalized density of free holes and trapped holes, respectively, by the right scale of the vertical axis.}
	\label{fig:Fit}
\end{figure*}

 \begin{threeparttable}[t]
\centering
\caption{Parameters fitted at each wavelength band and reference values. 
The acceptor density $N_A$ is fixed to $N_A=9\times 10^{14}~{\rm cm^{-3}}$. 
The errors show $1\sigma$ uncertainties. 
The derivation of the reference values is described in the text. 
NDF denotes the degrees of freedom. \label{tab:fitT}}
\begin{spacing}{1}
\begin{tabular}{lccccc}
\hline
band										& 6.45 $\mu$m	& 10.6 $\mu$m		&11.6 $\mu$ m		&15.1 $\mu$m 	& reference value	\\ \bhline{1.5pt}
$E_a$ [meV]									& $59\pm 4$ 	&$59\pm 4$			&$59\pm 7$			&$59\pm 6$		&	$\sim 60,~\sim 58.7$\tnote{*b}\\
$m_h/m_l$ 									& $7.5\pm0.5$		& $6.3\pm0.7$		& $6\pm1$			& $6\pm1$	& 	$\sim 5,~\sim7$\tnote{*c}\\
$a_0 ~[10^{-16} {\rm ~cm^{2}}]$					& $9\pm 1$ 		& $9\pm2$			& $8\pm3$			& $8\pm 3$ & 	--	\\
$b_0 [10^{-16}{\rm ~cm^{2}}]$					& 1 (fixed)\tnote{*d}		& $1.0\pm 0.6$		& $1.2\pm0.7$		& $1.0\pm 0.4$  &	--\\
$\sigma_{\rm trapped}[10^{-16}{\rm ~cm^{2}}]$	& $4.6\pm 0.1$	& $6.10\pm 0.07$	& $7.1\pm 0.1$		& $9.8\pm 0.1$  & 5--13\tnote{*e}\\ 
\hline
reduced-$\chi^2$								& 0.23			&  0.008\tnote{*f}		&	0.21				&	1.3		&		\\
NDF											& 4			& 3					& 3					& 3			&	\\ \hline
\end{tabular}
\end{spacing}
\begin{tablenotes}
{\footnotesize
\begin{spacing}{0}
\item[b] The values from the past CdTe studies \cite{molva1984,Ahmad2015}.
\item[c] The values from the past CdTe studies \cite{becker1989,Capek1973}.
\item[d] The $b_0$ value is fixed to secure convergence. 
\item[e] The predicted value based on the CdTe cryogenic absorption cross-section \cite{Capek1973}. 
\item[f] For the 10.6 $\mu$m band case, the reduced-$\chi^2$ is extremely small. 
This is probably because the uncertainty $\Delta\alpha$ is overestimated.
The uncertainty $\Delta\alpha$ is estimated based on the reproducibility of the measurement at 300 K before and after the cooling cycle. For measurement in the near temperature range, the uncertainty may be smaller than the estimated $\Delta\alpha$ because of the small displacement of the holder caused by a thermal shrink. 
\end{spacing}
} 
\end{tablenotes} 
\end{threeparttable}

Additionally, we compare derived physical parameters ($E_a$, $m_h/m_l$) to the previous studies to see if the fitted-parameter values are consistent with the previous measurements. 
We compare the best-fit acceptor energy $E_a$ to that of CdTe.  
The acceptor energy of Na-doped CdTe has been reported to be $\sim$ 60  meV using a Hall effect measurement \cite{Ahmad2015} and $\sim 58.7$ meV using a photoluminescence measurement \cite{molva1984}.  
The best-fitted acceptor energies ($E_a\sim 59$ meV)  are consistent at $1\sigma$ uncertainty level with CdTe previous studies. 
Following that, we compare the effective mass ratio $m_h/m_l$.
The best-fitted values of the effective mass ratio ($m_h/m_l=6$--7) are within the range of reported values by CdTe previous studies ($m_h/m_l\sim 7$\cite{Capek1973}, $\sim 5$\cite{becker1989}). 
These reference values based on previous studies are also listed in Table \ref{tab:fitT}.
Following the above comparison, the measured absorption coefficients and their temperature dependences are well-fitted with the absorption model by free/trapped holes and reasonable physical parameters ($E_a$, $m_h/m_l$). 

More so, we discuss the validity of the temperature dependence of the best-fit $\sigma_{\rm trapped}$. 
In this study, we assume that the trapped-hole cross-section $\sigma_{\rm trapped}$  is temperature independent as mentioned in the previous subsection. 
At $T<$50 K, the best-fit results show that the trapped-hole absorption is dominant. 
We conclude that the assumption of independence is reasonable, owing to the inability to see the temperature dependence of the $\alpha$ at $<50$ K.  

To validate the $\sigma_{\rm trapped}$ values, we compare the best-fit $\sigma_{\rm trapped}$ with a CdTe previous study, which is a similar substance used in the current study. 
\v{C}\'{a}pek et al. \cite{Capek1973} measured the $\alpha$ of $p$-type CdTe with $N_{\rm free}(300 {\rm~K})=3.1\times 10^{17}~ {\rm cm^{-3}}$ free-hole density. 
We assume that the trapped-hole density at 80 K is the same as the $N_{\rm free}(300~{\rm K})$ because the free-carrier density of CdTe and CdZnTe decreases with cooling \cite{Ahmad2015} as well as that of CdZnTe. 
As a result, the CdTe cross-section $\sigma_{\rm trapped,CdTe}$ can be estimated as follows:
\begin{eqnarray}
\sigma_{\rm trapped,CdTe}=\alpha(80~{\rm K})/N_{\rm trapped}(80~{\rm K})=\alpha(80~{\rm K})/N_{\rm free}(300~{\rm K}).
\end{eqnarray}
CdTe $\alpha$ values are taken from Fig. 3 of \v{C}\'{a}pek et al. \cite{Capek1973}. 
The $\sigma_{\rm trapped,CdTe}$ is calculated as $(5,~ 9,~ 10,~ 13)\times 10^{-16}~{\rm cm^{2}}$ at $\lambda=6.45,~10.6,~11.6,$ and $15.1~\mu$m, respectively. 
As a reference, the cross-sections $\sigma_{\rm trapped}$ derived for CdTe are also listed in Table \ref{tab:fitT}.
Also, because the best-fit $\sigma_{\rm trapped}$ in the current study is of the same order of magnitude as the $\sigma_{\rm trapped,CdTe}$, we conclude that the best-fit CdZnTe cross-section $\sigma_{\rm trapped}$ is within a reasonable range when compared to that derived for CdTe. 

In addition, we estimate the effect of the uncertainty of the acceptor density value $N_A$.
Based on Eq. 12, the acceptor density is assumed to be $N_A=9\times 10^{14}~{\rm cm^{-3}}$ as discussed in the previous subsection. 
From the uncertainty of the room-temperature resistivity, the uncertainty of the acceptor density is estimated as $\pm 5\times 10^{14}~{\rm cm^{-3}}$.
To estimate the effect of the uncertainty of the acceptor density, we perform the parameter fitting in the 10.6 $\mu$m band in the cases of $N_A=4\times 10^{14},~14\times 10^{14}~{\rm cm^{-3}}$. 
The fitting result is listed in Table \ref{tab:diffNa}. 
As shown in Table \ref{tab:diffNa}, the best-fit $E_a$ values are $\sim 60$ meV and the best-fit $m_h/m_l$ values are $\sim 6$ for all the $N_A$ cases. 
The best-fit $\sigma_{\rm trapped}$ values are of the same order of magnitude as that derived for CdTe above ($\sim9\times 10^{-16}~{\rm cm^2}$).  
We conclude that the uncertainty of the $N_A$ has minor effect on the estimation of the $E_a$ and $m_h/m_l$ but affect the best-fit value of $a_0,~b_0,~\sigma_{\rm trapped}$ by a factor. 

\begin{table}
\centering
\caption{
The best-fit result in the case of $N_A=(4,~9,~14)\times 10^{14}~{\rm cm^{-3}}$. The nominal $N_A$ value is $N_A=9\times 10^{14}~{\rm cm^{-3}}$. \label{tab:diffNa}
}
\begin{tabular}{lccc}\hline
											& case 1				& case 2	(nominal)		& case 3 \\ \bhline{1.5pt} 
$N_A[10^{14}{\rm cm^{-3}}]$ (fixed)					& 4				& 9				& 14	 \\ \hline
$E_a$ [meV]									& $63\pm5$			& $59\pm4$		& $56\pm4$	\\
$m_h/m_l$									& $6.5\pm0.7$		& $6.3\pm0.7$	& $6.2\pm0.7$	\\
$a_0~[10^{-16}~{\rm cm^{2}}]$					& $19\pm4$			& $9\pm 2$		& $6\pm2$	\\
$b_0~[10^{-16}~{\rm cm^{2}}]$					& $2.5\pm1$			& $1.0\pm 0.6$	&  $0.6 \pm0.4$  \\
$\sigma_{\rm trapped}~[10^{-16}~{\rm cm^{2}}]$	& $13.7\pm0.2$		& $6.10\pm0.07$	&  $3.92\pm0.05$ \\	\hline
\end{tabular}
\end{table}
 
In summary, the absorption coefficient due to free holes  decreases and becomes negligible at $T<50$ K, just as we initially predicted \cite{Sarugaku2017}. 
However, this is insufficient for explaining the temperature dependence of the measured absorption coefficient.
At cryogenic temperatures, the absorption coefficient due to trapped holes increases and becomes dominant at $T<50$ K. 
\v{C}\'{a}pek et al. \cite{Capek1973} showed the existence of the trapped-hole absorption in CdTe. 
Our study reveals that the trapped-hole absorption is observed also in CdZnTe. 
We note that \v{C}\'{a}pek et al. \cite{Capek1973} did not show the temperature dependence of the absorption by physical models, while our study reveals that the temperature dependence of the absorption coefficient is explained with the model from room temperature to low temperatures.

Moreover, we can predict the cryogenic absorption coefficient from the room-temperature resistivity based on the absorption model. 
According to the model, the cryogenic absorption is primarily caused by  carriers trapped at the impurity level. 
The cryogenic trapped-carrier density is roughly the same as the free-carrier density at room temperature because carriers trapped in the acceptor level at $T<50$ K are released to the valence band at $T\sim 300$ K. 
Also, the free-carrier density is known to be inversely proportional to the resistivity $\rho$.  
Hence, considering the above relationships, we derive an equation connecting cryogenic ($T<$50 K) trapped-carrier absorption and the room-temperature resistivity as follows: 
\begin{eqnarray}
\alpha_{\rm trapped}(T<50~{\rm K})&=&\sigma_{\rm trapped}N_{\rm trapped}(T<50~{\rm K}) \nonumber\\
		&\simeq& \sigma_{\rm trapped}N_{\rm free}(T=300~{\rm K})\propto 1/\rho(T=300~{\rm K}) \label{eq:predict}. 
\end{eqnarray} 
While trapped carriers are dominant absorbances, we can predict the cryogenic absorption coefficient from the room-temperature resistivity using the inversely proportional relation. 

Furthermore, we discuss the requirement of resistivity for application to a mid-infrared cryogenic immersion grating. 
The requirement of absorption coefficient of the immersion grating is $\alpha<0.01~{\rm cm^{-1}}$ \cite{Sarugaku2012}. 
Based on Eq. \ref{eq:predict} and our measurement results, to meet the $\alpha$ requirement,  CdZnTe must have  $\rho (T=300 ~{\rm K})> 10^4~{\rm \Omega cm}$. 
We also measure the $\alpha$ value of a high-resistivity CdZnTe in Appendix A to see if it meets the $\alpha$ requirement.
The high-resistivity CdZnTe results are consistent with the prediction based on the room-temperature resistivity (see Appendix A).

In addition, we anticipate that we can apply this model to other semiconductors to evaluate the cryogenic absorption coefficient (e.g., other semiconductor-type immersion gratings). 
This is because absorption causes in the model are common in general semiconductors \cite{Basu2003}, and are not unique to CdZnTe.  
 
\section*{CONCLUSIONS}
We measured the transmittance of CdZnTe substrates with two different thicknesses and estimated the absorption coefficient $\alpha$ in the $\lambda=6.45$, 10.6, 11.6, and 15.1 $\mu$m bands at $T=8.6$--300 K. 
At $T\sim 300$ K, the estimated $\alpha$ range is  $\alpha=0.3$--0.5 ${\rm cm^{-1}}$ which increases to $\alpha=0.4$--0.9 ${\rm cm^{-1}}$ at cryogenic temperatures ($T=8.6\pm 0.1$ K).  
Based on the physical absorption model, we propose that the dominant absorption cause at $T=$ 150--300 K is attributed to free holes: the dominant absorption cause at cryogenic temperature is attributed to trapped holes. 
Moreover, we discuss a method to predict the CdZnTe absorption coefficient at cryogenic temperature based on the room-temperature resistivity.

\section*{Conflict of Interest}
The authors declare that they have no conflict of interest.

\section*{Acknowledgments}We thank Mr. A. Noda and Dr. R. Hirano of JX Nippon Mining \& Metals Corporation for valuable comments. 
We appreciate members of the Laboratory of Infrared High-resolution spectroscopy (LIH) in Kyoto Sangyo University for informative advice on immersion grating and experiments for its material selections.  
This research is a part of conceptual design activity for the infrared astronomical space mission SPICA, which was a candidate for the ESA Cosmic Vision M5 and JAXA strategic L-class mission.
We appreciate the production of the holders and technical support by the Instrument Development Center in Nagoya University.
We are grateful for the technical assistance to construct the measurement system by Mr. R. Ito.
We appreciate Dr. J. Kwon and Mr. R. Doi for their supports in preliminary experiments prior to this study. 
H.M. is supported by the Advanced Leading Graduate Course for Photon Science (ALPS) of the University of Tokyo.

\section*{Appendix A: Absorption coefficient of the high-resistivity CdZnTe}
As discussed in the main text, the absorption coefficient of a higher-resistivity CdZnTe is lower than that of a low-resistivity CdZnTe.  
We measure the absorption coefficient of a high-resistivity-type CdZnTe to see if it has low absorption at cryogenic temperatures. 

First, we describe the high-resistivity CdZnTe sample and the measurement method.
As listed in Table \ref{tab:sampleHigh}, the samples in this appendix are $n$-type high-resistivity ($\rho>10^{10}~{\rm \Omega cm}$) CdZnTe substrates, while those in the main text are $p$-type low-resistivity ($\rho\sim 10^{2}~{\rm\Omega cm}$) CdZnTe substrates. 
We prepare thin-type and thick-type samples, highR-t1 and highR-t10.
The other conditions (surface roughness, Te-precipitate size, manufacturer) are the same as those described in the main text for samples. 
It is worth noting that the highR-t1 and highR-t10 are cut from different ingots. 
However, because they are manufactured in the same way, we assume that their absorption coefficients and reflectivities are similar.
The measurement method is the same as that described in the main text. 

Next, we describe the measurement result of the high-resistivity CdZnTe. 
Figure 5 shows the $\alpha$ of the high-resistivity CdZnTe at each temperature and wavelength. 
Also, because all the measured $\alpha$ are neither significantly above nor below zero, we set the 5$\sigma$ upper limits of the $\alpha<0.11~{\rm cm^{-1}}$.  
The $\alpha$ of the high-resistivity CdZnTe is lower than those of the low-resistivity one ($\alpha> 0.35 ~{\rm cm^{-1}}$) in the main text at all the measured wavelengths and temperatures. 

Following that, we compare our result with the previous result at room temperature. 
Kaji et al. \cite{Kaji2014} revealed that CdZnTe with lower resistivity absorbed more at room temperature. 
The room temperature $\alpha$ of the CdZnTe with a resistivity of $\rho=3.5 \times 10^{10}~{\rm \Omega cm}$ measured by them is less than $0.01~{\rm cm^{-1}}$in the wavelength range of $\lambda=$ 6.45--15.1 $\mu$m. 
The room-temperature absorption coefficient measured by them and that measured by us are consistent, although we can only restrict the $\alpha$'s upper limit. 
In addition, our result reveals that CdZnTe with lower resistivity has higher absorption even in the temperature range of 8.6--300 K, whereas they revealed this relation only at room temperature ($T\sim 295$ K). 

Next, we compare the high-resistivity CdZnTe with the $\alpha$ prediction based on room-temperature resistivity.
As discussed in the main text, the trapped-carrier absorption coefficient is inversely proportional to resistivity and mobility. 
According to the model, the main absorption cause is trapped-carrier absorption. 
Because the low-resistivity CdZnTe is $p$-type and the high-resistivity CdZnTe is $n$-type, we consider the absorption due to electrons trapped at the donor level. 
We assume that electron mobility is the same as in CdTe ($\sim 700~{\rm cm^2 V^{-1}sec^{-1}}$, \cite{wakaki2007}).
We also assume that trapped electron absorption follows the same order as trapped-hole absorption. 
Under these assumptions, the estimated absorption coefficient due to trapped electrons is negligible ($\alpha< 10^{-9}~{\rm cm^{-1}}$). 
As a result, we consider the lattice absorption $\alpha_{\rm lattice}$ and Te-precipitates attenuation $\alpha_{\rm Te}$, as dominant causes, which are ignored in the main text. 
As discussed in the main text, the sum of $\alpha_{\rm lattice}$ and $\alpha_{\rm Te}$ at $T$=8.6 K is considered to be less than that at $T=$ 300 K ($\alpha<0.006~{\rm cm^{-1}}$ at $\lambda=6.45$--15.1 $\mu$m).
Therefore, we can predict that the $\alpha$ of the high-resistivity CdZnTe at 8.6 K is $\alpha<0.006~{\rm cm^{-1}}$. 
Our result is consistent with this prediction, although the $\alpha$ has such a large uncertainty that we can only obtain the upper limit of the $\alpha$.  
A more precise measurement system is needed to reveal the temperature dependence of the $\alpha$ of high-resistivity CdZnTe because the $\alpha$ is lower than the uncertainty of our measurement system.

\begin{table*}
\caption{Properties of the high-resistivity CdZnTe single crystals}
\label{tab:sampleHigh}       \begin{tabular}{ccccccc}
\hline\noalign{\smallskip}
sample 	& incident-surface size 	&thickness  	& resistivity 				& conductivity 	& Te particle size	&	$Ra$ \\
		&  [mm$^2$]				&[mm] 		& [${\rm \Omega}$cm] 	&  type			&  [$\mu$m] 		&	[nm]\\ \bhline{1.5pt}
highR-t1		& 10$\times$10	& 0.79 		& $>10^{10}$		& $n$ 			& $<2$			& $<1$ \\
highR-t10 	&7$\times$7		& 9.94 		& $>10^{10}$		& $n$			& $<2$			& $<1$\\
\noalign{\smallskip}\hline
\end{tabular}
\end{table*} 

\begin{figure*}
	\includegraphics[width=129mm]{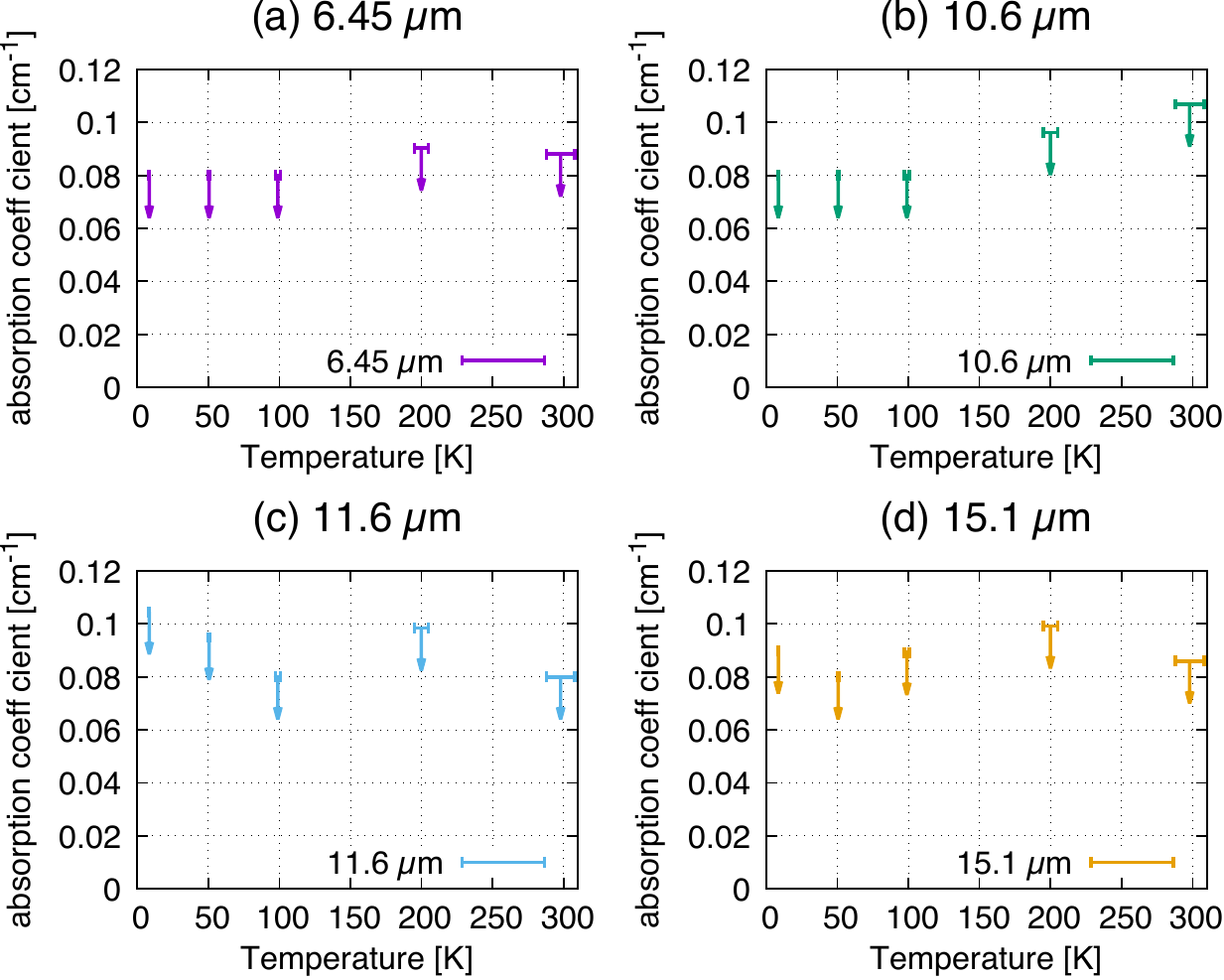}
\caption{
The upper limits of the $\alpha$  of the high-resistivity CdZnTe in the (a) 6.45, (b) 10.6, (c) 11.6, and (d) 15.1 $\mu$m bands.
Purple, green, sky-blue, and orange arrows show the 5$\sigma$ upper limit of the $\alpha$ in the 6.45, 10.6, 11.6, and 15.1 $\mu$m bands, respectively. }
	\label{fig:alphaH}
\end{figure*}

\section*{Appendix B: Sample holders}
Figure 6 shows the sample holders for thick and thin samples. 
To secure thermal contact, we screw the top of the thick-sample holders to the cold head surface with screws and hold them from the side with a copper plate and indium sheets. 
To monitor the temperature of the sample, we place a calibrated Cernox temperature sensor (CX-1030-SD-HT-1.4L; Lakeshore inc.) on the side of the sample. 
A temperature monitor (model 218, Lakeshore inc.) is used to monitor the sensor signal. 
On the other hand, the thin sample is clamped between the thin-sample holder and the holding plate. 
We screw the thin-sample holder to the cold head and insert indium sheets between the sample and the thin-sample holder to secure thermal contact, 
To monitor the sample temperature, we use the same temperature sensor used in the thick-sample case on the holder.

\begin{figure*}
	\includegraphics[width=129mm]{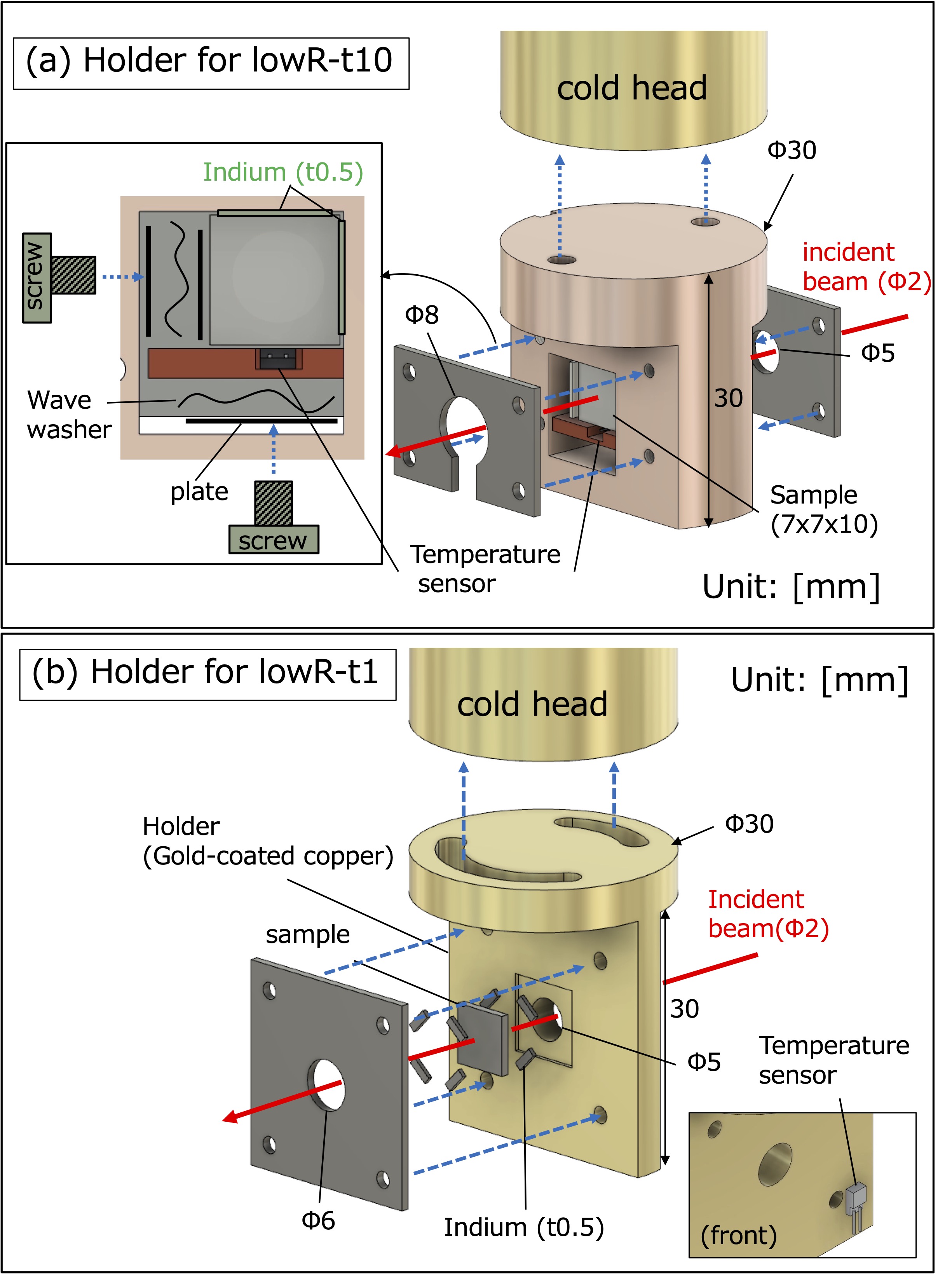}
\caption{Two types of sample holders: (a) holder for lowR-t10 sample and (b) holder for lowR-t1 sample. 
Holders are screwed to the cold head of the cryostat. 
Red arrows show the beam path. 
Blue dashed arrows indicate screwed-in points. 
Dimensions are in millimeters.
Details are described in the text. 
}
	\label{fig:holders}
\end{figure*}

\end{document}